\documentclass[preprint,prl,aps,tightenlines,floatfix,showpacs]{revtex4}

\usepackage{graphics}
\usepackage{bm}
\usepackage{amsmath}
\usepackage{amssymb}

\begin{document}

  \title{Electron scattering form factors from exotic nuclei}
  \author{S. Karataglidis$^{(a)}$}
  \email{S.Karataglidis@ru.ac.za}
  \author{K. Amos$^{(b)}$}
  \email{amos@physics.unimelb.edu.au}
  \affiliation{$^{(a)}$  Department of  Physics and  Electronics, P.O.
    Box 94, Rhodes University, Grahamstown, 6140, South Africa}
  \affiliation{$^{(b)}$ School of Physics, University of Melbourne,
  Victoria, 3010, Australia.}
  \date{\today}
  
  \begin{abstract}
    The  elastic electron  scattering form  factors,  longitudinal and
    transverse, from the  He and Li isotopes and  from $^8$B have been
    studied. Large space shell  model functions have been assumed. The
    precise distribution  of the neutron  excess has little  effect on
    the form factors of the isotopes though there is a mass dependence
    in  the  charge  densities.  However,  the  form  factors  of  the
    proton-rich  nucleus,  $^8$B,  are  significantly changed  by  the
    presence of the proton halo.
  \end{abstract}
  \maketitle
  
  The form factors from  elastic electron scattering off exotic nuclei
  is  of  topical  interest  given  the proposed  construction  of  an
  electron-ion collider  at GSI \cite{Si05} and the  effort to measure
  electron  scattering form factors  of exotic  nuclei at  RIKEN using
  Self-Confining  Radioactive Ion  Targets (SCRITs)  \cite{Su05}. Such
  experiments   will  provide   direct  information   of   the  charge
  distributions in  exotic nuclei  that lie outside  of the  valley of
  stability  and whose  structures may  have  skins or  halos for  the
  excess  nucleons. Also, through  appropriate kinematic  selection of
  collision events, those facilities  may obtain transverse as well as
  longitudinal   form  factors.  Taken   together  with   analyses  of
  nucleon-nucleus ($NA$) scattering  data which provide information on
  the  matter densities~\cite{Am00},  analyses of  these  form factors
  should make  feasible a significant  map of the densities  of exotic
  nuclei.

  Electron   scattering  from  exotic   nuclei  has   been  considered
  previously by  Antonov \textit{et  al.} \cite{An05}, for  both light
  and heavy systems. In  particular, for scattering from light nuclei,
  they used a Helm model  to obtain longitudinal form factors. Therein
  densities   obtained   from   no-core   shell   model   calculations
  \cite{Ka97,Ka00} also  were used. With  the He isotopes,  they found
  that the charge  density of $^6$He differed from  that of $^4$He but
  that of  $^8$He was not  significantly different to that  of $^6$He.
  They  also  noted that  the  proton  density  does extend  far  with
  increasing neutron number. However, they did not make any conclusion
  as to the  role of the neutron halo character  of $^6$He. That role,
  of the halo in elastic electron scattering, was sought more recently
  by Bertulani~\cite{Be06}. He also used the Helm model and found that
  the  contribution from  the neutron  halo itself  was insignificant.
  However, for  proton-rich nuclei, he  observed that the  proton halo
  did so,  and significantly. That is  to be expected  given the large
  extension of  the charge  density over  that of a  skin such  a halo
  engenders.

  As those recent works~\cite{An05,Be06} used the Helm model, only the
  longitudinal form factors could  be considered. Essentially they are
  the Fourier transforms  of the charge density. It  is the purpose of
  this letter to consider both the longitudinal and transverse elastic
  electron scattering form factors for the He and Li isotopes from the
  valley  of stability  to the  drip lines  and to  use a  large space
  no-core shell  model to  define the required  structure information.
  For comparison, we have studied  the form factors of the proton-halo
  nucleus, $^8$B.

  The microscopic model for  electron scattering specified by deForest
  and   Walecka~\cite{Fo66}    and   by   Karataglidis,    Halse   and
  Amos~\cite{Ka95}  and based  on the  shell model,  has been  used to
  calculate the  form factors. Using  the notation of the  latter, the
  form factors  for electron  scattering between nuclear  states $J_i$
  and $J_f$ involving angular momentum transfer $J$ are expressed as
  \begin{equation}
    \left| F^{\eta}_J(q) \right|^2 = \frac{1}{2J_i+1} \left(
    \frac{4\pi}{Z^2} \right) \left| \left\langle J_f \left\|
    T^{\eta}_J(q) \right\| J_i \right\rangle \right|^2,
  \end{equation}
  where  $\eta$  selects   the  type,  i.e.  longitudinal,  transverse
  electric, or  transverse magnetic. Assuming  one-body operators, the
  reduced matrix elements may be expressed in the form,
  \begin{equation}
    \left\langle J_f \left\| T^{\eta}_J(q) \right\| J_i \right\rangle
    = \frac{1}{\sqrt{2J+1}}\text{Tr}(SM),
  \end{equation}
  where  $S$   is  the   matrix  of  one-body   transition  densities,
  $S_{j_1j_2J}$, defined as
  \begin{equation}
   S_{j_1j_2J} = \left\langle J_f \left\| \left[ a^{\dag}_{j_2}
      \times \tilde{a}_{j_1} \right]^J \right\| J_i \right\rangle.
  \end{equation}
  $M$  denotes the  matrix elements  of the  one-body  longitudinal or
  transverse electromagnetic operators  for each allowed particle-hole
  excitation  ($j_2$-$j_1^{-1}$).  Bare  operators  are used  for  the
  results presented herein,  and explicit meson-exchange-current (MEC)
  effects are ignored. However,  MEC have been incorporated implicitly
  in the  transverse electric form  factors in the long-wave  limit by
  using Siegert's  theorem \cite{Fr85}. That serves  to introduce into
  the transverse  electric form factor  an explicit dependence  on the
  charge density,  through the use of the  continuity equation. Hence,
  any  effect on  the charge  density from  a halo  would  affect both
  longitudinal and transverse electric form factors.

  Within the framework of the shell model, the halo has been specified
  by using  Woods-Saxon (WS) single-particle (SP)  wave functions with
  the binding  energies of the  orbits occupied by the  valence (halo)
  nucleons set to  the appropriate single-nucleon separation energies.
  Doing so allows for the  extension of the nucleon density consistent
  with there being a halo~\cite{Ka00}. Conversely, when the density is
  specified using harmonic  oscillator (HO) wave functions, consistent
  with  the  shell  model,  a  neutron skin  results.  Hence,  results
  obtained with  WS and HO SP  wave functions are denoted  as halo and
  non-halo,  respectively. This distinction  is not  new. It  has been
  used to  explain the anomalously  large $B(E1)$ value  for $^{11}$Be
  with  simple shell  model wave  functions~\cite{Mi83}. Also,  such a
  distinction  has allowed  identification  of the  halo attribute  in
  $^6$He  in analyses of  elastic and  inelastic scattering  of $^6$He
  from hydrogen~\cite{La01}

  We first discuss  the form factors for electron  scattering from the
  even-even  He   isotopes.  The   shell  model  wave   functions  for
  $^{4,6,8}$He  were  obtained  from a  complete  $(0+2+4)\hbar\omega$
  shell model using  the $G$ matrix shell model  interactions of Zheng
  \textit{et  al.} \cite{Zh95}. The  halo in  $^6$He was  specified by
  using  WS wave  functions with  the  SP energies  of the  $0p$-shell
  neutrons  set to  the single  neutron separation  energy  of 1.8~MeV
  \cite{Ka00}. An  oscillator parameter of 1.8~fm was  used to specify
  the HO SP wave functions.

\begin{figure}
    \scalebox{0.4}{\includegraphics*{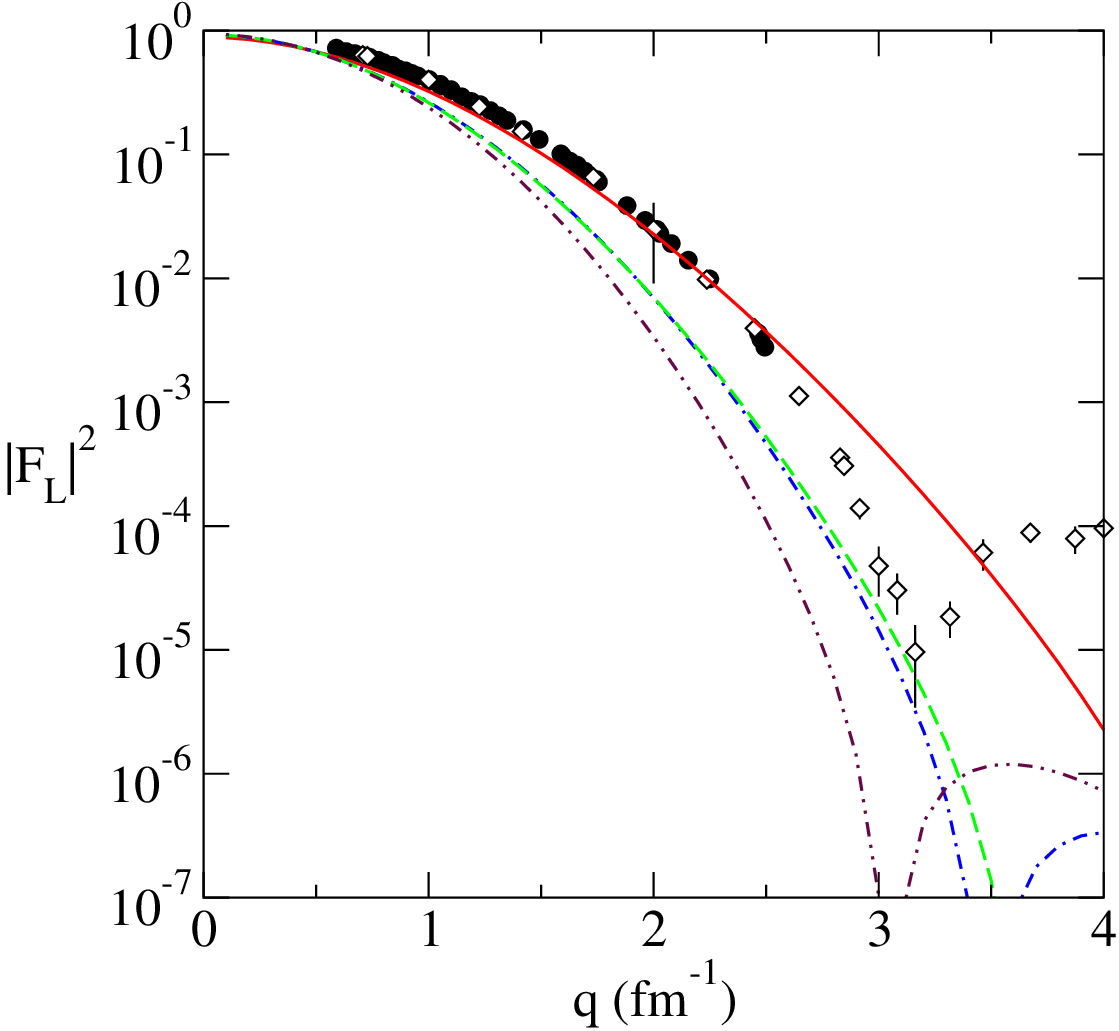}}
    \caption{\label{helium}    (Color   online.)    Elastic   electron
    scattering form  factors for  $^4$He (solid line),  $^6$He (dashed
    line,  halo;  dot-dashed   line,  non-halo),  and  $^8$He  (double
    dot-dashed line). The data for the $^4$He form factor are those of
    McCarthy \textit{et al.} \cite{Mc77}.}
  \end{figure}

  The longitudinal  elastic electron scattering form  factors from the
  He isotopes are displayed in Fig.~\ref{helium}.
  Therein, the form factors for  the scattering from $^4$He and $^8$He
  are portrayed  by the solid  and double-dot-dashed lines,  while the
  results of the calculations made  using the halo and non-halo models
  of $^6$He are given by the dashed and dot-dashed lines respectively.
  The data  for the $^4$He form  factor are those of  McCarthy {\em et
  al.} \cite{Mc77}. The comparison of the $^4$He form factor with data
  is quite  good up  to $2.5$~fm$^{-1}$. This  is consistent  with the
  predicted charge radius of 1.71~fm as compared to the measured value
  of  $1.671 \pm 0.014$~fm  \cite{Ot85}. The  addition of  neutrons to
  form $^6$He and $^8$He pull the charge density out and thus the form
  factors decrease  with momentum transfer. Note that  there is little
  difference between the results of the halo and non-halo calculations
  for  the $^6$He form  factor. The  form factor  is not  dependent on
  detailed properties of the neutron  halo. It is only the presence of
  the  extra  2  neutrons  that   causes  the  change  to  the  proton
  distribution.  The Helm  model  results of  Antonov \textit{et  al.}
  \cite{An05}  for the  $^6$He form  factor generally  agree  with the
  present   ones.  However,   Antonov   \textit{et  al.}   \cite{An05}
  under-predict the $^4$He form factor as compared to the transform of
  the  experimental  charge  density,  and somewhat  over-predict  the
  $^8$He form factor.  It would seem that, while  the charge densities
  of the  isotopes are essentially the  same, that they  have used for
  $^4$He is too extended.

\begin{figure}
    \scalebox{0.4}{\includegraphics*{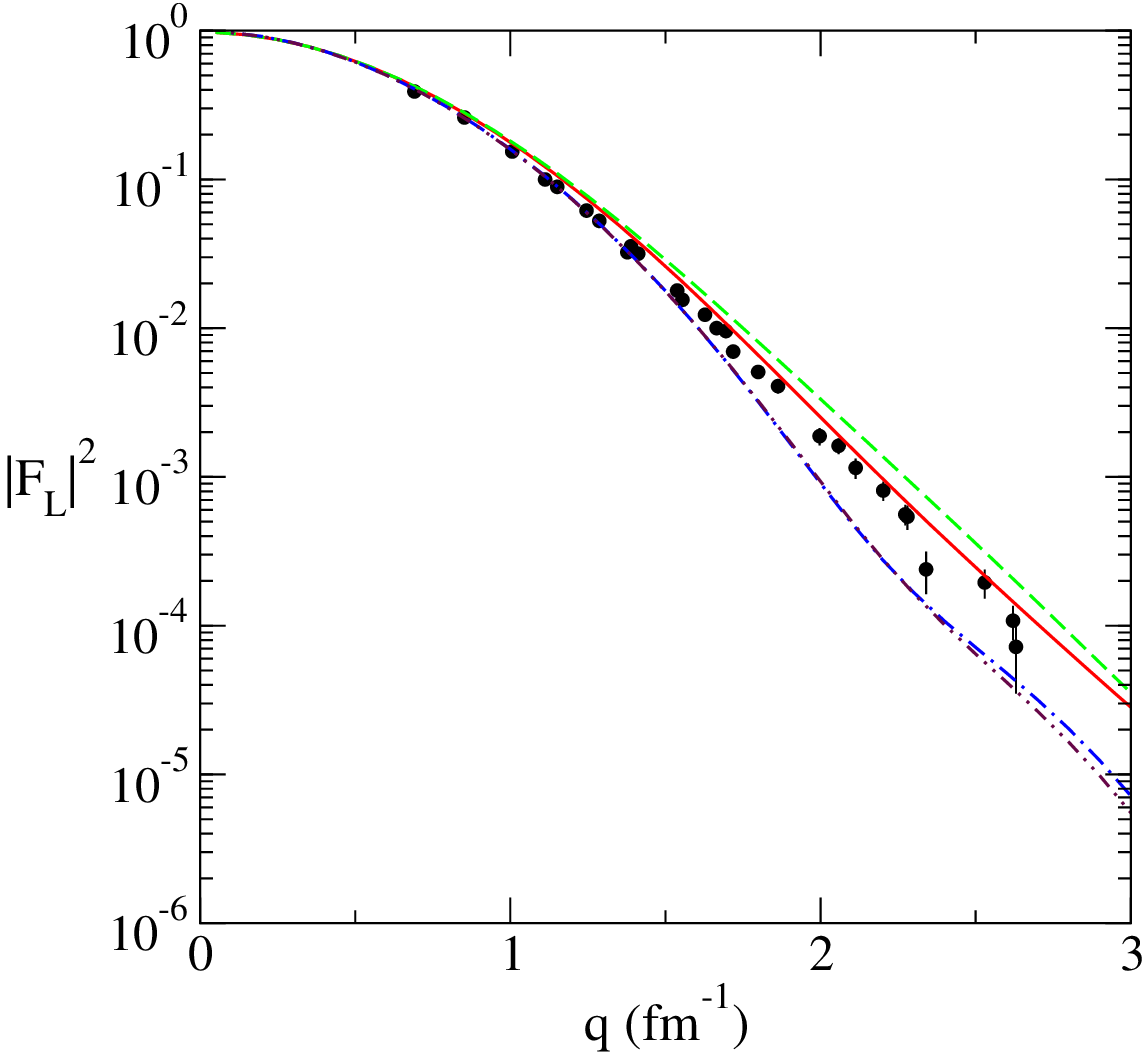}}
    \caption{\label{lilong}  (Color online.) The  longitudinal elastic
      scattering form  factors for $^7$Li, $^9$Li,  and $^{11}$Li. The
      results  of  the  calculations  made  for  $^7$Li,  $^9$Li,  and
      $^{11}$Li  are portrayed  by the  solid, dashed,  and dot-dashed
      (non-halo) and double-dot-dashed (halo)  lines. The data for the
      $^7$Li    longitudinal    form    factor    are    from    Refs.
      \cite{Su67,Li89}.}
  \end{figure}

  The longitudinal elastic scattering form factors for $^7$Li, $^9$Li,
  and $^{11}$Li are displayed in Fig.~\ref{lilong}.
  The level of  agreement between the results of  our calculations for
  $^7$Li with the  data of Suelzle \textit{et al.}  \cite{Su67} and of
  Lichtenstadt \textit{et al.} \cite{Li89} is quite good. The addition
  of  two  neutrons  to  $^7$Li   does  not  change  the  form  factor
  substantially and so the charge density for $^9$Li is little changed
  from that  for $^7$Li.  But a noticeable  change is observed  in the
  form factor  for $^{11}$Li. It  decreases faster than the  other two
  with increasing momentum transfer. As with $^6$He, this extension of
  the charge  density does not come  about with the  halo specifics in
  this nucleus. Rather, it is due  only to the coupling of the 4 extra
  neutrons to the  $^7$Li core. A similar conclusion  has been reached
  by Bertulani \cite{Be06}.

\begin{figure}
    \scalebox{0.4}{\includegraphics*{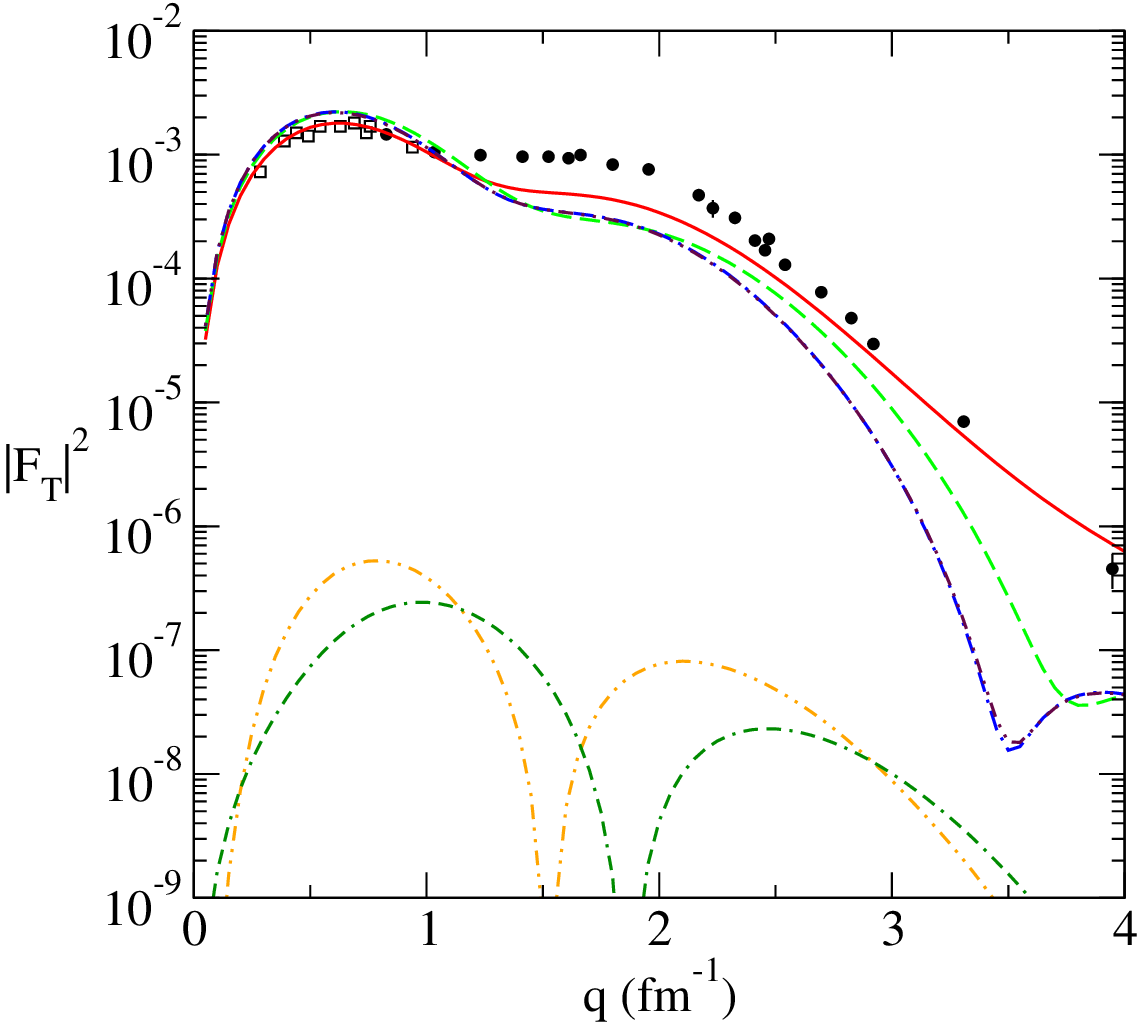}}
    \caption{\label{litran}  (Color online.) As  for Fig.~\ref{lilong}
    but  for   the  transverse  form   factors.  The  data   are  from
    Refs.~\cite{Li89,Ni71}. The halo and nonhalo neutron components of
    the  M1   form  factor  from   $^{11}$Li  are  portrayed   by  the
    double-dot-dashed  and dot-double-dashed  lines,  respectively, at
    the bottom of the figure.}
  \end{figure}

  The transverse  elastic scattering form factors  for $^7$Li, $^9$Li,
  and $^{11}$Li are displayed in Fig.~\ref{litran}.
  These form  factors are  dominated by $M1$  effects at  low momentum
  transfer values,  to $\sim 1$~fm$^{-1}$,  but are dominated  by $E2$
  contributions  above that value~\cite{Ka97}.  The $M1$  form factors
  for all three nuclei are  quite similar. It is dominated entirely by
  the proton contribution. Such is expected as that form factor is not
  influenced   by  changes   in  the   charge  density.   The  neutron
  contribution  to  the $^{11}$Li  $M1$  form  factor,  also shown  in
  Fig.~\ref{litran}, is influenced by  the neutron halo but is several
  orders of magnitude below the proton component. The $E2$ form factor
  shows significant decreases in value at large momentum transfers for
  both $^9$Li and $^{11}$Li from  the reference $^7$Li form factor. As
  with the  longitudinal $^{11}$Li form  factor, that decrease  is due
  only to the addition of the 4 neutrons about the $^7$Li core and not
  due to the halo characteristic.

\begin{figure}
    \scalebox{0.4}{\includegraphics*{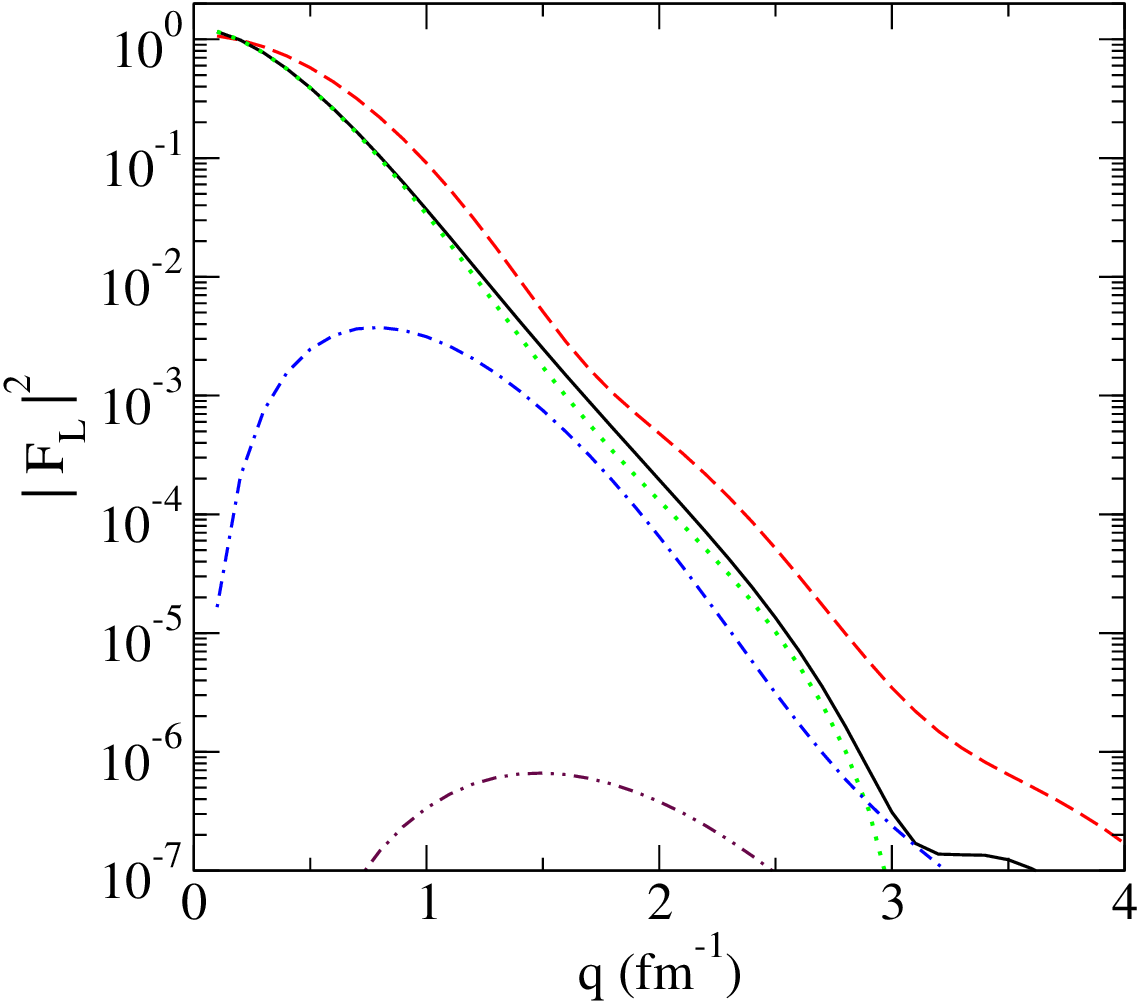}}
    \caption{\label{b8long}   (Color  online.)   Longitudinal  elastic
    scattering  form factor  for $^8$B.  The halo  and  non-halo model
    results are portrayed by the solid and dashed lines, respectively.
    The $C0$,  $C2$, and $C4$ components  of the halo  form factor are
    given  by  the dotted,  dot-dashed,  and double-dot-dashed  lines,
    respectively.}
  \end{figure}

  As  a comparison  to the  neutron  halos, we  now consider  electron
  scattering  from a  proton halo  nucleus, namely  $^8$B.  A complete
  $(0+2+4)\hbar\omega$  shell   model  using  the   Zheng  interaction
  \cite{Zh95}  was  used  to  obtain  the  OBDME.  The  single  proton
  separation energy  from $^8$B is 137~keV \cite{TUNL04}  and the halo
  is specified by using WS  functions for $0p$-shell protons with that
  binding energy. The  non-halo specification uses the same  set of WS
  functions as for $^8$He. The prediction for the longitudinal elastic
  electron  scattering   form  factor  from  $^8$B   is  presented  in
  Fig.~\ref{b8long}.
  This  form   factor  is  dominated  by  the   $C0$  component  below
  1~fm$^{-1}$  while the $C2$  contribution becomes  significant above
  that  value. The  $C4$ component  is negligible.  The effect  of the
  proton halo, in this case, is  very evident with the decrease in the
  form  factor with  momentum transfer  reflecting the  more extensive
  charge distribution.

\begin{figure}
    \scalebox{0.4}{\includegraphics*{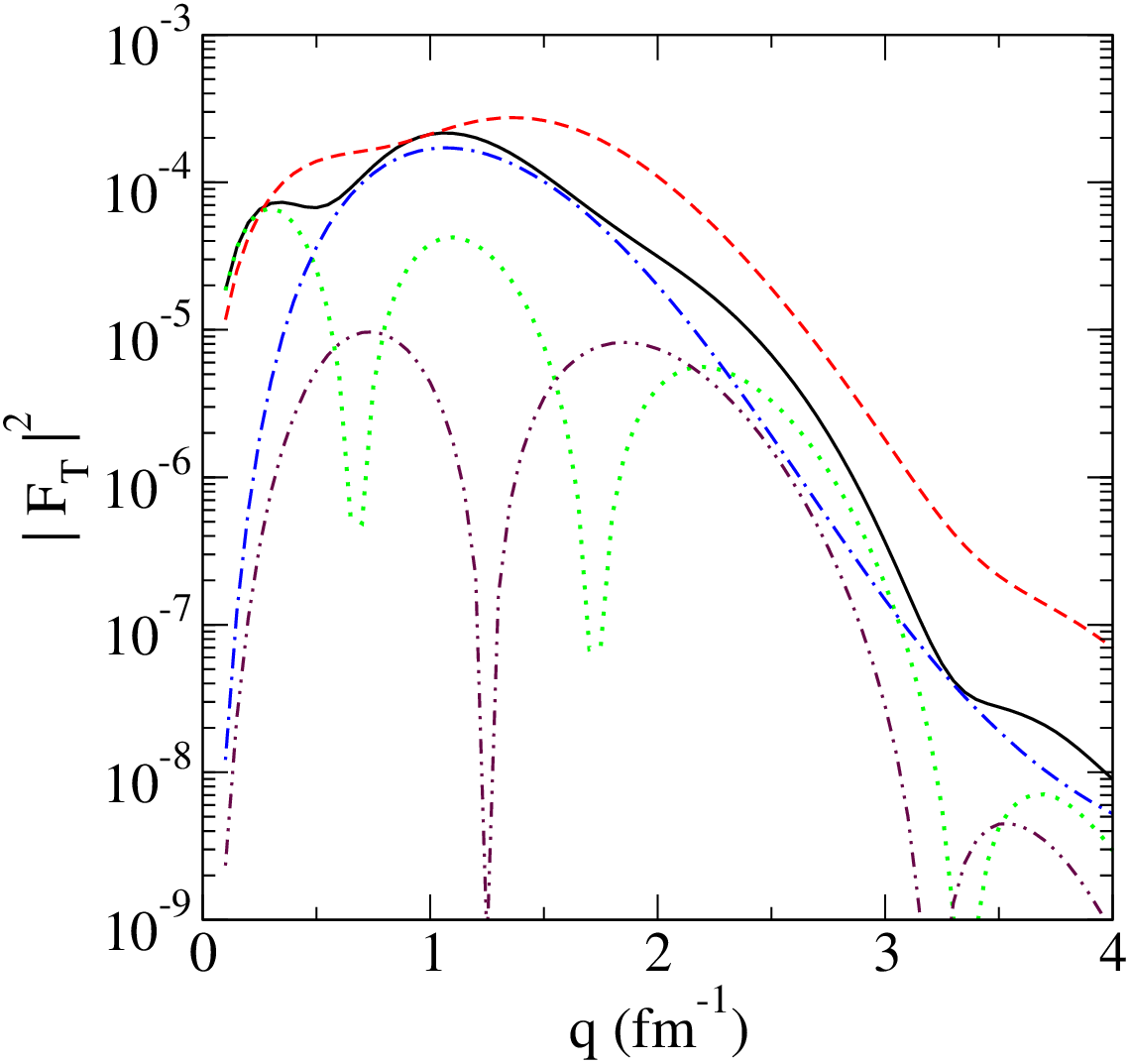}}
    \caption{\label{b8tran}   (Color   online.)   Transverse   elastic
    scattering form factor  for $^8$B. The form factors  from the halo
    and non-halo calculations are given by the solid and dashed lines.
    The $M1$,  $E2$, and  $M3$ form factors  are given by  the dotted,
    dot-dashed, and double-dot-dashed lines, respectively.}
  \end{figure}

  The transverse elastic scattering form  factor for $^8$B is shown in
  Fig.~\ref{b8tran}.
  The effect of the proton halo  on this form factor for $^8$B is just
  as dramatic as that on the longitudinal one. The change in this case
  is a significant  decrease in all components of  the form factor. As
  it is  dominated by  the $E2$ component,  the total  transverse form
  factor  decreases markedly  due to  the  presence of  the halo.  The
  magnetic  components are  also significantly  modified by  the proton
  halo, although  that manifests itself only  in a change  at very low
  momentum transfer where the $M1$ component dominates.

  Results have been presented for the elastic electron scattering form
  factors from exotic  nuclei to elicit details of  the effects of the
  extended  nucleon  matter   distributions  on  the  charge  density.
  Significantly,  the neutron  halo character  does not  influence the
  electron scattering form  factors, as was shown in  reference to the
  form factors of $^6$He and $^{11}$Li. Instead, the charge density is
  extended  naturally only  through the  addition of  neutrons  to the
  stable isotopes. That will  occur irrespective of whether the exotic
  nucleus has a neutron halo or a neutron skin.

  The  situation   for  proton   halos  is  entirely   different.  The
  longitudinal  and transverse  electron scattering  form  factors for
  $^8$B are significantly reduced compared to the standard shell model
  result. This is consistent with  the extension of the charge density
  due to the proton halo. The transverse form factors portrays similar
  behavior due to  the dominance of the $E2$  component, which is also
  affected by the proton halo.

  With the prospect of electron scattering form factors being measured
  soon   with   SCRIT  \cite{Su05}   or   the  electron-ion   collider
  \cite{Si05},  it is hoped  that investigations  of the  proton halos
  will be possible. Transverse form factors should also be measured as
  such measurements  will also be  possible in the  collider. Analysis
  with  complementary RIB-hydrogen scattering  data then  suggests the
  possibility  of a  significant mapping  of the  densities  of exotic
  nuclei so  providing detailed microscopic tests  of structure models
  in use.
  
  \bibliography{draft3-ehalo}

\end{document}